\documentclass[12pt]{article}
\begin{document}

\begin{center}
{\large Michel H\'enon's first research article:\\[2mm]
An improved calculation of the perturbation of stellar velocities}\\[6mm]
{\it Fathi Namouni\\[1mm]
Universit\'e de Nice, CNRS, Observatoire de la C\^ote d'Azur}
\end{center}

\vspace{10mm}

{\it Abstract: Fifteen years after the discovery of dynamical friction by Chandrasekhar, Michel H\'enon attempts to solve the longstanding problem of the divergence of the friction suffered by the perturber and caused by the most distant cluster stars. His solution laid the foundation to the current understanding of dynamical friction as a non-local transitory force.}

\vspace{5mm}

Michel H\'enon started a PhD thesis at the Institut d'Astrophysique de Paris in 1957 that addressed the modeling  of star clusters dynamics. Chief among the processes that enter the relaxation of such systems is dynamical friction. The phenomenon was discovered fifteen years earlier by Subrahmanyan Chandrasekhar in an influential paper that set the foundation of the statistical description of gravitational systems. Chandrasekhar had recognized that the theory of random flights implied a linear time increase of the square of velocity dispersion. He reasoned that the presence of a friction process is necessary in order to make the Maxwellian probability distribution function invariant. Chandrasekhar then derived a Fokker-Planck-type equation for the probability distribution function whose friction and diffusion coefficients were obtained from the simple model of pairwise encounters: orbital deflection of a test star by a cluster field star was assumed to be isolated and complete. However, the averaging of the friction felt by the test star over the  cluster  stars diverges as the size of the spatial integration domain. In {\it Principles of stellar Dynamics (1942)}, Chandrasekhar attributed the appearance of such divergence to two causes: the first is the overestimation of the contribution to friction from stars farther away than the average interstellar distance from the test star. The second cause is the inadequacy of the  simple two-body problem for describing the dynamics of a multi-body problem when the distance to the test star is comparable to the interstellar distance. He estimated that the cutoff distance that must be applied to the diverging integral should be within a factor of 2 or 3 of the interstellar distance, and added that  {\it `it would be difficult to estimate $D_{\tt max}$ [the cutoff distance] more closely than this without going into a considerable amount of detailed calculation.'} Enter PhD student Michel H\'enon. In an analytical article titled in French `Un calcul am\'elior\'e des perturbations des vitesses stellaires (Probl\`eme du temps  de relaxation)' published in 1958 in Annales d'Astrophysique, volume 21, page 186,  H\'enon demonstrated that the divergence of dynamical friction is an artefact of the assumption of full binary encounters and that the causes invoked by Chandrasekhar were largely irrelevant. He did so by calculating with  great physical accuracy the diffusion and friction coefficients of Chandrasekhar's theory. The improvement that H\'enon  refers to in his title is the introduction of the star's travel time and the realization that during that time the test star's orbit  will suffer complete hyperbolic deflection from nearby stars and incomplete deflection from faraway stars. To derive the effect of distant perturbers, he suggested the use of classical celestial mechanics' perturbation theory. As in practice the test star's traveled  distance is larger that its interaction distance, H\'enon chose an arbitrary  distance scale $l$ somewhere between the latter two that allows him to apply full hyperbolic deflection for impact parameters smaller than $l$ and slight deflection for impact parameters ranging from $l$ to {infinity}. The diffusion and friction coefficients thus estimated turned out to be converging and independent of the arbitrary length $l$. The analytical integrals also allowed him to examine how the test star's neighborhoods contributed to friction and diffusion. He showed for instance that intermediate stars do not contribute to the velocity dispersion along a star's motion. H\'enon's most important result however is the appearance of time in the Coulomb logarithm giving the correct spatial cutoff of Chandrasekhar's theory as the product of the velocity dispersion and the time over which the system is observed. For a single star that crosses the medium, the cutoff is the traveled distance. Dynamical friction is therefore shown to be a non-local transitory process. Years later, these results were rederived through different methods by Prigogine and Severne, Physica (1966) volume 32, page1376 and Ostriker and Davidsen, Astrophysical Journal (1968) volume 151, page 679. The importance of the transitory nature of dynamical friction  may be illustrated by the resolution of the once-paradox of gravitational friction in a infinite homogeneous gaseous medium felt by a perturber moving in uniform rectilinear motion. Whereas the force for supersonic motion was shown to be inversely proportional to the square of the perturber's velocity akin to Chandrasekhar's formula, it was argued that when motion is subsonic, friction disappears. The reason is that the front-back symmetry of the density perturbation generated by the moving body cancels the back reaction of the medium on the perturber leading to the unphysical situation of a supersonic perturber that decelerates by a friction  whose amplitude increases with decreasing velocity until Mach 1 is reached when the force suddenly disappears. The issue was resolved by Ostriker in the Astrophysical Journal (1999) volume 513, page 251 who adopted a transitory approach similar to H\'enon's theory for collisionless systems to study the time-dependent disturbance that the perturber introduces in the background medium. Dynamical friction was found to be finite for subsonic motion and proportional to the logarithm of the traveled distance for supersonic motion exactly as in H\'enon's work. Viewed as the first research article of a PhD student, H\'enon's work on dynamical friction and its completion of Chandrasekhar's theory showed the promise of what was to come from one astronomy's great scientists. Viewed in a more contemporary context and despite an obvious lack of citations in the literature, H\'enon's first research article has influenced and continues to indirectly  influence various areas of astrophysics from the evolution of planets around massive stars to the growth of super massive black holes by mergers in colliding galaxies.

\end{document}